\documentclass{mpe_report}

\usepackage{psfig,graphicx,epsfig}
\usepackage{color}
\usepackage{amsmath,amssymb,epic,eepic,array}
\usepackage{natbib}

\begin{document}

\pagenumbering{arabic}
\setcounter{page}{177}

\renewcommand{\FirstPageOfPaper }{177}\renewcommand{\LastPageOfPaper }{180}\title{Asymmetric neutrino emission in quark matter and pulsar kicks}
\author{I. Sagert and J. Schaffner-Bielich}  
\institute{Institute for Theoretical Physics/Astrophysics, J. W. Goethe University, D-60438 Frankfurt am Main, Germany}
\maketitle 
\begin{abstract}
The puzzling phenomenon of pulsar kicks, i.e. the observed large escape velocities of pulsars out of supernova remnants, is examined for compact stars with a strange quark matter core. The direct Urca process in quark matter is studied in the presence of a strong magnetic field. Conditions for an asymmetric emission of the produced neutrinos are worked out in detail, giving constraints on the temperature, the strength of the magnetic field and the electron chemical potential in the quark matter core. In addition, the neutrino mean free paths for quark matter and a possible hadronic mantle are considered. 
\end{abstract}

\section{Introduction}

In the course of a supernova explosion and the following cooling stage of the proto neutron star neutrinos carry away around 99\% of the energy which is produced in the core collapse. This binding energy $E_B$ of the iron core is approximately:
\begin{eqnarray}
E_B&\sim&3G_N M_c^2/(5R_c)
\nonumber\\
&\sim&3\cdot 10^{53}\mathrm{erg}\left[\frac{M}{1.4 M_\odot}\right]^2\left[\frac{10\mathrm{km}}{R}\right]
\label{neutrino_en}
\end{eqnarray}
where $G_N$ is the gravitational constant, $M_c$ and $R_c$ the mass and the radius of the iron core, respectively and $M_\odot$ the mass of the sun. The neutrinos are mainly produced during the core collapse by electron capture as well as during the Helmholtz-Kelvin cooling stage of the proto-neutron star. Due to the large densities exotic matter can occur in the interior of the star in the form of kaon- or pion condensates as well as hyperons or even deconfined strange quark matter composed of up, down and strange quarks \citep{Weber05}. In the latter case the neutron star can have either just a quark matter core (hybrid star) or be entirely composed of strange quark matter (strange or selfbound star, respectively). An overview on strange quark matter in compact stars is given by \citet{Schaffner-Bielich05}. Similar to electrons which form Cooper pairs in metals, quarks can form pairs with net colour charge which is referred to as colour superconductivity. In the interior of a neutron star quarks are present in three different flavours and three colours giving a large variety of pairing possibilities and therefore different colour superconducting phases. However, for the largest densities it is assumed that quarks with all colours and flavours will pair to the so-called colour-flavour locked phase (CFL), where electric charge neutrality is given by the strange quarks and therefore the electron chemical potential is zero. \citet{Ruester04} showed the appearance of the metallic CFL phase (mCFL), also referred to as modified CFL, for temperatures larger than $\sim$ 10 MeV. Here the electron chemical potential can become very large and chemical potentials of the quarks are in the range of 400 MeV. 

\section{Pulsar kicks} 

Pulsars, which are highly magnetized, rotating neutron stars, are observed to have proper motions up to 1600 kilometres per second \citep{Hobbs}. As their progenitor stars move with 10-50 km/s this raises the question about the origin of the high pulsar velocities, which are denoted in general as pulsar kicks. Though measurements of space velocities have large error bars, \citet{Brisken} found the highest directly measured velocity for a pulsar is 1080$\pm 100$ km/s. Another interesting fact concerning the pulsar proper motions is the alignment between the velocity vector and the rotational axis of the Crab and the Vela pulsar as well as PSR B0656+14. These observations give strong indications for a connection between the pulsar kick direction and the rotational axis and the magnetic field, respectively \citep{Johnston}. However, till today not one of the proposed acceleration mechanisms could successfully explain the full spectrum of the measured proper motions within the standard model . An overview of the most common approaches can be found in \citet{Wang} or \citet{Lai}.

\section{Pulsar kicks by anisotropic neutrino emission}
In this work we introduce an acceleration mechanism due to asymmetric neutrino emission from the direct quark Urca process in the interior of proto-neutron stars. A neutron star which moves with a velocity of 1000 km/s has a kinetic energy of $\sim 10^{49}$ erg. Hence just a small asymmetry in the total neutrino emission with the energy of (\ref{neutrino_en}) could accelerate the neutron star up to velocities of 1000 km/s. The neutrinos stem from the direct quark Urca process which is the main neutron star cooling mechanism in the presence of a quark matter core:
\begin{eqnarray}
d \longrightarrow u + e^{-} + \bar{\nu_e},\hspace{2mm}u + e^{-} \longrightarrow n + \nu_e
\end{eqnarray}
The asymmetry in the neutrino emission arises due to a strong magnetic field which can align the electron spin opposite to the magnetic field direction. Due to parity violation the polarisation of the neutrino spin direction will fix the neutrino and anti-neutrino momenta. As a result all neutrinos and antineutrinos leave the star in one direction accelerating it reversely. To check whether this acceleration can work we have to ensure that the required magnetic field is smaller than the maximum one of $B_{max} \sim 10^{18}$ G for a neutron star of 1.4 solar masses and a radius of 10 km. Furthermore we have to calculate the size of the kick velocity which is produced by the neutrino emission. As a third point we have to check the neutrino mean free paths in the neutron star interior. According to the no-go theorem by \citet{Vilenkin} neutrinos which are produced in a statistical and thermal equilibrium can loose their polarisation due to high interaction rates which means in our case the loss of the acceleration mechanism. Consequently we have to ensure that the neutrino mean free paths are large enough.

\section{Kick velocity calculations}

The specific heat capacity per volume $c_v$ gives the ability of matter to store heat for changing temperatures. It can be calculated by the neutrino emissivity $\epsilon_{q\beta}$ from the direct quark Urca process with:
\begin{eqnarray}
c_v dT=- \epsilon_{q\beta} dt.
\label{hc_eps}
\end{eqnarray}
For the neutron star interior composed of normal quark matter the heat capacity will be dominated by quarks. Consequently we can neglect the heat capacity of the electrons whereas for quarks we also have to take into account strong interactions to first order in the coupling constant $\alpha_s = $g$^2/(4\pi)\simeq 0.5$ (at energies of interest here) which gives
\begin{eqnarray}
c_q=9\mu^2T\left(1-\frac{2 \alpha_s}{\pi}\right)=A\mu^2T
\label{cap_q}
\end{eqnarray}
with the degeneracy factor g=18 (from spin, colour and flavour, as we are considering up and down and strange quarks to be present). Due to momentum conservation the kick velocity can be determined with the neutrino luminosity $L = 4\pi R^3/3 \cdot\epsilon_{q\beta}$ to be:
\begin{eqnarray}
dv&=&\frac{\chi L}{M_{ns}}dt=-\frac{4}{3}\pi R^3\frac{\chi}{M_{ns}}c_q dT\\
v&=&\frac{2}{3}\pi R^3\frac{\chi}{M_{ns}}A{\mu_q}^2T^2,
\label{velocity}\\
&\sim& 38\frac{\mathrm{km}}{\mathrm{s}}\left(\frac{\mu_q}{400\mathrm{MeV}}\frac{T_0}{10^{10}\mathrm{K}}\right)^2\left(\frac{R}{10\mathrm{km}}\right)^3\frac{1.4M_\odot}{M_{\mathrm{ns}}}\chi
\label{velocity2}
\end{eqnarray}
where $T$ is the initial temperature. We can neglect the final value due to the fast temperature decrease in the proto neutron star cooling evolution (see \cite{Pons}). For an initial temperature of $T_0\sim 5\cdot 10^{10}$ K $\sim 5$ MeV a kick of 1000 km/s is possible for a quark phase radius of $R \sim 10$ km which would be typical for a strange star. 

\section{Polarisation of electrons - Landau Levels}

\begin{figure}
\centerline{\psfig{file=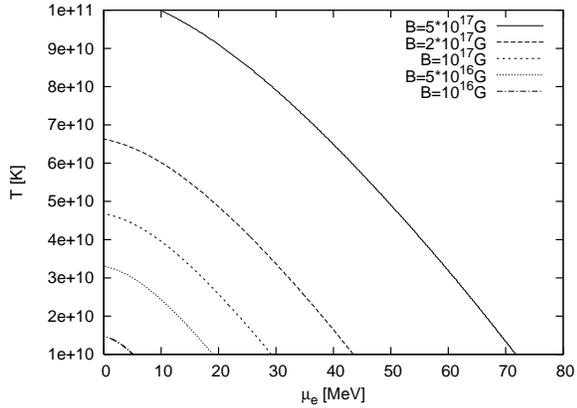,width=5.5cm,angle=270} }
\caption{Temperature, electron chemical potential and magnetic field strength B which are needed to fully polarize the electrons.
\label{pol}}
\end{figure}

Electrons which move in an external magnetic field have orbital motions on the plane perpendicular to the magnetic field direction, whereas the radius decreases with increasing magnetic field strength $B$. For 
\begin{eqnarray}
B>B_{crit}\sim\frac{m_e c^2}{e\hbar}\sim4.4\cdot10^{13}G.
\label{mag_field}
\end{eqnarray}
the radius even comes in the range of the de Broglie wavelength of the electron and the system becomes quantised, whereas the quantised energy of the electron has the form:
\begin{eqnarray}
E^2&=&{m_e}^2+{p_z}^2+2eB\eta.
\label{energy_n}
\end{eqnarray}
The magnetic field points in the positive z-direction and $p_z$ is the electron momentum along the magnetic field. The quantisation manifest itself in discrete electron orbits, the so-called Landau levels, which are labelled in equation (\ref{energy_n}) by $\eta$ and are defined by their quantum number $\nu$ and the electron spin $s$ in the following way:   
\begin{eqnarray}
\eta = \nu + \frac{1}{2} + s \mbox{ and } 
\label{landaulevel}
s & = & \left\{ \begin{array}{ccccc}+\frac{1}{2}\quad\mbox{ for spin-up}\\-\frac{1}{2}\quad\mbox{   for spin-down} \end{array}\right.
\label{spin} 
\end{eqnarray}
For a sufficient large magnetic field strength all electrons are sitting in the lowest Landau level with $\eta$ = 0. The electron number densities $n_{+}$ and $n_{-}$ in denote the electrons with spin parallel or anti-parallel to the magnetic field direction, respectively. From equation (\ref{spin}) one sees that the lowest Landau level with $\eta=0$ can only be occupied if $\nu = 0$ and $s=-\frac{1}{2}$. Consequently the electrons situated here are spin polarised with $s=-\frac{1}{2}$.
Assuming that we can treat the electrons in the neutron star interior as an ideal gas of fermions we can write the number densities of electrons with spin-up and spin-down as:
\begin{eqnarray}
n_\mp&=&\frac{geB}{(2\pi)^2}\sum_{0}^{\nu_{max\mp}}\int_0^\infty f(E)dp_z.
\label{number_den1}
\end{eqnarray}
where $p_z$ is the momentum in the direction of the magnetic field, $f(E)$ is the Fermi distribution function and $\nu_{max\mp}$ are the maximum Landau quantum numbers whereas $\eta = \nu_-=\nu_+ +1$. Having calculated both number densities, $n_-$ and $n_+$, one can finally determine the polarisation
\begin{eqnarray}
\chi=\frac{n_- - n_+}{n_- + n_+}.
\label{chi}
\end{eqnarray}
Figure \ref{pol} shows the constraints on the magnetic field strength to fully polarise the electrons for a given electron chemical potential $\mu_e$ and temperature $T$, showing that the necessary magnetic field for full polarisation increases with $\mu_e$ and $T$.
As seen in the previous section the temperatures which are required to accelerate the neutron star to velocities of 1000 km/s are in the range of $T \sim 5\cdot 10^{10}$ K. From Figure \ref{pol} we find that the required magnetic fields are in the range of $B \sim 10^{17}$ G to achieve full polarisation for different electron chemical potentials. These magnetic fields seem to be very high compared to the observed ones of approximately $B \sim 10^{12}$ G. Nevertheless, one should take into account that these high magnetic fields are just required in the core of the neutron star and can decrease on the way to the surface. Finally so-called magnetars presumably have large polar magnetic fields of $B > 10^{14}$ G at the surface.

\section{Neutrino mean free paths}
\label{neutrino_mean}

Up till now we found that neutrinos stemming from a neutron star interior with temperatures around 5 MeV can accelerate the neutron star up to 1000 km/s. Furthermore we discussed the strength of the magnetic field coming to the conclusion that for the required temperatures of $\sim 5$ MeV a magnetic field with $B \sim 10^{17}$ G could fully polarise the electron spin. Next we consider the neutrino mean free paths. For the neutron star interior we have to consider at least four neutrino interaction processes with the medium: the absorption of neutrinos in quark matter ($d+\nu_e\longrightarrow u+e^-$) and neutron matter ($n + n + \nu_e\longrightarrow n + p + e^-$) as well as the scattering processes ($q+\nu\longrightarrow q+\nu$ and $n+\nu\longrightarrow n+\nu$) for both interiors. We will give these mean free paths following closely the work of \citet{Iwamoto81}. Yet, we have to keep in mind that Iwamoto did not consider magnetic field effects and assumed degenerate electrons, that is, the electron chemical potential was supposed to be much larger than the temperature. The neutrino mean free paths taken from \citet{Iwamoto81} are summarised in table \ref{irina_tab} and on the first sight seem to be quite large. However, inserting a temperature of 5 MeV, which is required for kicks around 1000 km/s, we find that the mean free paths decrease drastically to the range of about 100 m. \citet{Reddy02} found a large neutrino mean free path in CFL quark matter of more than 10 km for $T$ = 5 MeV. Unfortunately the interaction with massless Goldstone bosons decreases the mean free path again to the range of 100 m.
\begin{table}
\begin{tabular}{|l||l||l|}
\hline
\textsc{Medium} & \textsc{Process} &\textsc{Mean free path}\\
\hline
\hline
Quark matter&Absorption&l$_{abs}\sim 25 \mbox{km}\frac{MeV^3}{T^2\mu_e}$\\
Quark matter&Scattering&l$_{scat}\sim 92 \mbox{km}\left(\frac{MeV}{T}\right)^3$\\
Neutron matter&Absorption&l$_{abs}\sim 17 \mbox{km}\left(\frac{MeV}{T}\right)^3$\\
Neutron matter&Scattering&l$_{scat}\sim 244 \mbox{km}\left(\frac{MeV}{T}\right)^4$\\
\hline
\end{tabular}
\caption{Neutrino mean free paths for absorption and scattering in quark matter as well as in neutron matter for $E_\nu=3 T, \mu_q=400$ MeV, $\alpha_s=0.5$.}
\label{irina_tab}
\end{table}
Neutrino absorption and neutrino scattering will occur in the interior of the star. Consequently, one can combine the neutrino mean free paths for absorption and scattering in a total mean free path for quark matter as well as neutron matter by:
\begin{eqnarray}
\frac{1}{l_{total}} = \frac{1}{l_{abs}} + \frac{1}{l_{scatt}}.
\end{eqnarray}
\begin{figure}
\centerline{\psfig{file=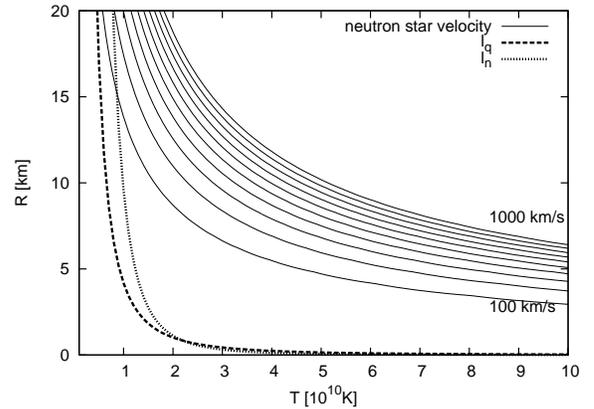,width=5.5cm,angle=270} }
\caption{Neutrino mean free paths in quark matter and neutron matter for $E_\nu=3 T$, $\mu_q = 400$  MeV, $\alpha_s=0.5$ and $\mu_e=10$ MeV with kick velocities for $M_{ns}=1.4M_\odot$, $\chi = 1$.
\label{vel_mfp}}
\end{figure}
The combination of the radius and temperature dependence of the kick velocity with the total neutrino mean free paths for quark and neutron matter are plotted in Figure \ref{vel_mfp}. It shows different neutron star velocities for a given initial temperature $T$ and a quark phase radius $R$. The lowest curve corresponds to a neutron star velocity of 100 km/s whereas the velocity increases for each curve in steps of 100 km/s, i.e. the highest curve represents neutron stars with $v$ = 1000 km/s. The stars are assumed to have a total mass of $1.4 M_\odot$. For the required temperature of 5 MeV the mean free paths are very small, thus the neutrinos will interact with the neutron star medium on their way to the surface and isotropise. Consequently, the accelerating mechanism will be washed out. 

\section{Possible solutions for small neutrino mean free path}

\begin{figure}
\centerline{\psfig{file=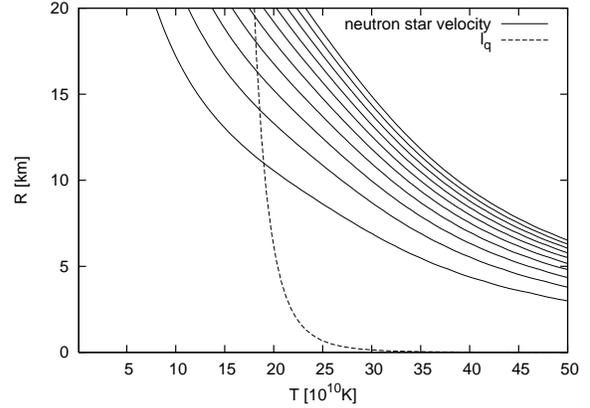,width=5.5cm,angle=270} }
\caption{Kick velocities and neutrino mean free path in quark matter for the CFL quark phase with $\alpha_s = 0.5$, $\mu_q = 400$ MeV, $\mu_e = 100$ MeV and $\Delta = 100$ MeV.
\label{mfp_gap}}
\end{figure}

In the previous sections we saw that the neutrino mean free paths in normal strange quark matter are too short to assure free streaming neutrinos. We will study now the neutrino emission from quark matter in the metallic CFL phase as it provides high electron chemical potentials as well as exponentially suppressed neutrino quark interactions. We find that while the pairing between the quarks enhances the neutrino mean free paths exponentially with $e^{\Delta/T}$, where $\Delta$ is the gap energy it also decreases the quark heat capacity by the same factor:
\begin{eqnarray}
c_q=A{\mu_q}^2Te^{-\Delta(T)/T}.
\end{eqnarray}
For large values of the gap the quark heat capacity is lowered drastically. Consequently, the electron heat capacity, which was considered before to be negligible becomes now crucial. With the electron chemical potential $\mu_e$ it can be written as: 
\begin{eqnarray}
c_e=\frac{{\mu_e}^2T}{2}+\frac{7}{30}g\pi^2T^3.
\end{eqnarray}
Implementing $c_q$ and $c_e$ in the velocity calculation in (\ref{velocity}) we plot the kick velocities in Figure \ref{mfp_gap} assuming:
\begin{eqnarray}
\Delta = 150 \mbox{ MeV} \mbox{ and }\mu_e = 100 \mbox{ MeV}.
\nonumber
\label{parameters}
\end{eqnarray}
The kick velocity and the mean free paths are plotted in Figure \ref{mfp_gap} in the same way as for Figure \ref{vel_mfp} using $\alpha_s = 0.5$ and $\mu_q = 400$ MeV.
Comparing now Figures \ref{vel_mfp} and \ref{mfp_gap} one finds that the electron heat capacity is lower than the one from ungapped quark matter. For temperatures in the range of $2\cdot 10^{11}$ K - 3 $\cdot 10^{11}$ K (ignoring the neutrino mean free path) velocities of 100 to 1000 kilometres per second can be reached for a quark phase radius of approximately 10 km. For increasing temperatures the gap suppression factor $e^{-\Delta/T}$ diminishes and the contribution from quark matter becomes important again. As all neutrino-quark interactions are suppressed the neutrino mean free path is greatly enlarged. However, at the same time all velocity curves are shifted to higher temperatures. Consequently, just low velocities can be reached for free streaming neutrinos. As can be seen in Figure \ref{mfp_gap} a quark phase radius of $\sim 11$ km is required to accelerate the star to 100 km/s.

\section{Conclusions}

We calculated for a fixed neutron star mass of $M_{ns}=1.4M_\odot$, a quark chemical potential of 400 MeV and fully spin polarised electrons the pulsar kick velocity in dependence of the quark phase temperature and its radius. For an initial temperature of $T_0 \sim 5\cdot 10^{10}$ K and magnetic fields of $B \sim 10^{17}$ G a kick of 1000 km/s is possible for $R \sim 10$ km which would be typically for a strange star. For these neutron star properties the neutrino mean free paths are in the range of 100 m and therefore too small to prevent neutrinos from interacting and isotropising. Assuming quark matter in the CFL phase we get for $\Delta = 150$ MeV, $\mu_e = 100$ MeV, $\mu_q = 400$ MeV and $T \sim 1.8\cdot 10^{11}$ K a kick velocity of 100 km/s within a quark phase radius of 11 km for free streaming neutrinos i.e. ignoring effects from Goldstone bosons lowering the $\nu$ mean free path.
\vskip 0.4cm

\bibliographystyle{aa}
\bibliography{all}

            \clearpage

\end{document}